\documentstyle[12pt,epsf]{article}
\newcommand{\nsigma}{\mbox{\boldmath $\sigma$}}
\newcommand{\ntau}{\mbox{\boldmath $\tau$}}
\begin{document}
\title{\Large \bf Short-range Correlations in a CBF description of
closed-shell nuclei.\footnote{Presented by G.Co' at the Second
Workshop on Electromagnetically Induced Two--Nucleon Knockout,
Gent, May 17--20, 1995.}\\ }
\author{F.Arias de Saavedra$\,^1$, G.Co'$\,^3$, A.Fabrocini$\,^4$, \\
 S.Fantoni$\,^5$, I.E.Lagaris$\,^2$,  A.M.Lallena$\,^1$\\ \\
{\small 1) Departamento de Fisica Moderna, Universidad de Granada,} \\
     {\small \sl  Granada, Spain}  \\
{\small 2) Department of Physics, University of Ioannina,} \\
{\small \sl Ioannina, Greece }  \\
{\small 3) Dipartimento di Fisica, Universit\`a di Lecce and} \\
{\small INFN, Sezione di Lecce,}\\
{\small \sl Lecce, Italy }  \\
{\small 4) Dipartimento di Fisica, Universit\`a di Pisa and} \\
{\small INFN, Sezione di Pisa,}\\
{\small \sl  Pisa, Italy }  \\
{\small 5) Interdisciplinary Laboratory for Advanced Studies (ILAS) and} \\
{\small INFN, Sezione di Trieste,}\\
{\small \sl Trieste, Italy } }

\date{\mbox{ }}
\maketitle

\begin{abstract}
The Correlated Basis Function theory (CBF) provides a theoretical
framework to treat on the same ground mean--field and short--range
correlations. We present, in this report, some recent results obtained
using the CBF to describe the ground state properties of finite nuclear
systems.  Furthermore we show some results for the excited state obtained
with a simplified model based on the CBF theory
\end{abstract}
\newpage
\section{Introduction}
In the description of a many--body system, the word {\bf
correlation} indicates the fact that the state of each particle of
the system depends on the presence of the other particles.

In an infinite system of interacting particles, this definition of
correlation can be exactly translated saying that {\sl
anything beyond the Fermi--gas model is a correlation}. For a
finite set of particles the phrase {\sl anything beyond the shell
model is a correlation}, does not translate the definition of
correlation given above since the mean--filed is already
correlating the particles of the system, localising their presence
around a specific point of the space.

A proper description of a finite many--body system should treat on
the same ground the correlation generated by the mean--field
and those produced by the residual interaction among the particles.

The Correlated Basis Function theory (CBF) provides a theoretical
framework for this consistent treatment. The CBF has been applied
successfully to both few--body systems and nuclear matter (for a
review see ref. \cite {pan90}).

In recent years, we have started a work aimed to apply the CBF
to the description of medium and heavy nuclei \cite{co92,co94}. In
this report we shall present some recent results of this work.

\section{CBF and FHNC theories}
The CBF is based upon the variational principle
\begin{equation}
\label{ritz}
\delta E[\Psi]=\delta \frac {<\Psi | H | \Psi>} {<\Psi | \Psi>} =0,
\end {equation}
which corresponds exactly to the Schr\"odinger equation
if the variation is performed considering the full Hilbert space of
many--body wave functions, or in other words, if no limitations on the
structure of the many--body wave function $|\Psi>$ are enforced.

On the other hand, one solves
the variational principle instead of the Schr\"odinger equation
because one wants to work in a limited sub--space of the Hilbert space.
For example, the assumption that $|\Psi>$ is a Slater determinant
of single particle wave functions leads to the Hartree--Fock equations.

The assumption on the structure of $|\Psi>$ used in the CBF is:
\begin{equation}
\label{function}
\Psi(1,2,...,A)=F(1,2,...,A)\Phi(1,2,...,A)
\end{equation}
where $\Phi$ is a Slater determinant and $F$ is a many--body
correlation function defined as a symmetrized product of two-body correlation
functions:
\begin{equation}
\label{corr}
F(1,...,A)=
{\cal S} \prod_{i<j}
[ \sum_{n=1}^{M} f^{(n)} (r_{ij})O^{(n)}(i,j) ]
\end{equation}
In the above equation $\cal S$ represents the symmetrizer operator
and the two--body correlation functions have been expressed in a most
general form, where the state dependence, given by the operators $O^n$,
is, in general, the same of the hamiltonian \cite{pan79}. This correlation
operator contains only explicit two--body correlations. More general forms,
mainly used in light nuclei, may also include three--body correlations.

The variational procedure consists
 in performing the variation on the correlation
function and on the set of single particle wave functions in order to
find the minimum of the energy functional of eq. \ref{ritz}. This
requires the evaluation of complicated multidimensional integrals.
The most direct approach to tackle the problem is the numerical
evaluation performed with Monte Carlo technology. In this case the full
procedure is called Variational Monte Carlo (VMC) calculation.
This brute force method is, unfortunately, not suitable to describe
medium and heavy nuclei since the number of spin--isospin states to be
sampled becomes enormous (for the $^{40}$Ca nucleus is of the order of
the Avogadro number).

Since we are interested in
the description of medium and heavy nuclei we used a different technique:
 the cluster expansion. This technique is better
illustrated   when the correlation function is
purely scalar and commutes with the nuclear potential, $V$. In this case the
mean value of $V$ can be written as:
\begin{equation}
<\Psi|V|\Psi>= <\Phi|FVF|\Phi>=<\Phi|VF^2|\Phi>=
<\Phi|V\prod_{i<j}^A(f^2(r_{ij}))|\Phi>
\end{equation}
where $r_{ij}$ represents the distance between the $i$ and $j$
nucleons. The last equality of the previous equation follows from the
orthonormality of the single particle wave functions.
We definine a function $h$ as:
\begin{equation}
f^2(r)=1+h(r).
\end{equation}
The importance of this function lies in the fact that it is appreciably
different from zero only in a small region of $r$, and it can be used as
an expansion parameter.
The mean value of the potential becomes:
\begin{eqnarray}
\label{cluster}
\nonumber
&&\,\,<\Psi|V|\Psi>=<\Phi|V\prod_{i<j}^A(1+h(r_{ij}))|\Phi> \\
&=&<\Phi|V(1+h(r_{12}))(1+h(r_{13}))
...(1+h(r_{23}))...|\Phi>.
\end{eqnarray}

The cluster structure clearly shows up in the above equation.

Let's consider the term obtained retaining only the $1$'s in the
expansion of $F^2$. We obtain:
\begin{equation}
\label{clust}
<\Phi|V|\Phi>
\end{equation}
which is the mean value of the potential between the uncorrelated
states.
If $V$ is a two--body potential
\begin{equation}
\label{potential}
V= \frac {1}{2} \sum_{i<j}^Av(r_{ij}) ,
\end{equation}
the sum of all the terms linear in $h$ generates
two--body cluster diagrams
\begin{equation}
\frac {A(A-1)}{2} <\Phi|v(r_{12})h(r_{12})|\Phi> ,
\end{equation}
and three-- or four--body cluster diagrams of the form:
\begin{equation}
\frac {A(A-1)}{2} <\Phi|v(r_{12})\sum_{i<j}^Ah(r_{ij})|\Phi> .
\end{equation}
The three--body diagrams are obtained when
either $i$ or $j$ is equal to $1$ or $2$ and the other index is
different, while in the four--body diagrams both $i$ and $j$ are different
from $1$ and $2$.
Summing all the terms quadratic in the $h$ functions,
\begin{equation}
<\Phi|V\sum h(r_{ij})h(r_{lm})|\Phi> ,
\end{equation}
three--, four--, five-- and six--body cluster diagrams are obtained.

The procedure continues considering all the terms given by eq.
\ref{cluster}.
A similar expansion can be derived for the kinetic
energy mean value.

The analysis of the cluster expansion is done in a more efficient
way using the so--called Mayer diagrams \cite{may40}. From this
analysis it turns out that the mean value of the hamiltonian between
the two correlated states can be expressed simply in terms of two
classes of diagrams: the nodal diagrams and the elementary, or
bridge, diagrams (see references \cite{pan79,ros81,fab86} for more
details).

The Fermi Hypernetted Chain (FHNC) provides a set of integral
equations which allows one to sum up all the nodal diagrams. There is
not closed form to do the same for the elementary diagrams. They should
be calculated one by one like in ordinary perturbation theories.

A usual approximation in this kind of calculations, the so--called FHNC/0
approximation, consists in neglecting the elementary diagrams. It is
worth  to stress here that
the FHNC/0 calculations differ from the Monte
Carlo ones only by the fact that the elementary diagrams are not
considered. The relevance of these diagrams has been studied
in nuclear matter where it has been found that their contribution to the
energy is small \cite{pan90,pan79,wir95}.
This is not the case for other many-body
systems, like liquid helium \cite{ari92}.

Our first goal was to test the validity of FHNC theory against VMC
calculations \cite{bue92}. For this reason
we have performed calculations in model nuclei
using scalar correlation functions and semi-realistic interactions.
Our model nuclei are composed by
protons and neutrons having the same radial
single--particle wave functions, their angular momentum coupling is done
in the $l-s$ scheme, and the Coulomb interaction is switched off.

The comparison between our results and the VMC ones are
presented in tab.1 where, in addition to the binding energies $<E>$, the
contribution of the Majorana $<v_M>$ and Wigner $<v_W>$
terms of the potential and that of the
kinetic energy $<T>$ are shown.

The calculations have been performed using the
Brink and Boeker B1 interaction \cite{bri67} and with harmonic
oscillator single particle wave functions.

The FHNC results are presented in the rows labelled FHNC/0. We
observe the fact that in $^{16}$O
 the binding energies differ from the VMC results by a factor
of about the 10\%. More important was the fact that the spin sum rule,
\begin{equation}
S_{\sigma}= \frac {1}{3A} \int \, d^3r_1 d^3_2
<\Psi|\sum_{i\neq j} \delta( {\bf r_1-r_i})
\delta ({\bf r_2-r_j}) \nsigma_i \cdot \nsigma_j
|\Psi>=-1,
\end{equation}
which should hold for the spin saturated systems we have studied,
was not satisfied by a 10\% factor.

Since the only difference between FHNC/0 and VMC calculations is the
absence in the former calculations of the elementary diagrams, we
attributed to them the discrepancy between
the two results.
For this reason we have performed other calculations inserting the lowest
order elementary diagram. The results of these calculations are
presented in tab.1 by the rows labelled as FHNC-1.
We found that the inclusion of the
elementary diagram was practically affecting
only the Majorana part of the potential
energy, and therefore we expected big changes in the spin sum rule. In
effect the spin sum rule resulted to be satisfied at the 1\% level.

These results, in spite of the importance of some classes of
elementary diagrams, gave us confidence about the validity of
our approach.

The next step of our work consisted in applying the
formalism to real nuclei, with  different radial wave
functions for protons and neutrons, angular momentum coupling
in a $j-j$ scheme, and with Coulomb interaction.
In this new situation we found that
the $j-j$ coupling is generating, in the nuclei
with unsaturated $l$ levels, a new statistical correlation.

In tab.2. we present the results of calculations performed keeping
fixed the Woods-Saxon mean--field which generates the single
particle wave functions and changing the correlation function
in order to minimise the binding energy. The interaction used
is the S3 force of Afnan and Tang \cite{afn68} extended to consider
the odd channels as described in ref. \cite{co92}.

The results of the column $F1$ have been obtained
using the same mean--field potential for protons and
neutrons (a Woods-Saxon well without spin--orbit term),
switching off the Coulomb interaction and  the new
statistical correlation.
The effect of this new correlation can be seen comparing
the results of column $F1$ with those column $F2$. The only difference
with the calculation of the $F1$ column is just the inclusion of the new
correlation terms. The effect of this new correlation is present only in
the potential energy term of the $^{12}$C, $^{48}$Ca and $^{208}$Pb nuclei
which have some unsaturated $l$ shell, and it is rather small.

The $F3$ column presents the results obtained
adding the Coulomb interaction in the hamiltonian.
The contribution of the nuclear interaction ($V$) to the
binding energy is about the same for $^{40}$Ca, $^{48}$Ca and $^{208}$Pb.
This nuclei are sufficiently large to allow the nuclear interaction to
{\sl saturate}. On the other hand, the contribution of the Coulomb
interaction ($V_c$), because of the infinite range of the force,
increases like the number of the protons pairs.

The $F4$ column shows the results obtained inserting the
Coulomb potential also in the mean--field. We have used the
potential generated by a homogeneously charged sphere. Other
choices did not make any relevant difference.

Instead of
performing a full minimisation of the energy functional changing both the
mean--field and the correlation function, we took from the
literature Woods--Saxon potentials fixed in order to
reproduce single particle energies around the Fermi surface,
and with these mean--fields we minimised the
binding energies finding the optimal correlation functions for
both the B1 and S3 interactions. The results of these
calculations are presented in tab.3 and they show
the same characteristics
of those of tab.2.
The nuclear interaction {\sl saturates}
for the heavier nuclei while the contribution of
the Coulomb interaction increases with the number of protons.

It is interesting to compare the results obtained by the
FHNC/0 calculations (rows $E$) with those obtained including
the elementary diagram (rows $E/4$). The contribution of the
elementary diagram is noticeable in the lighter nuclei. In
$^{12}C$ is of the order of the 10\% and even more. As soon as the
number of particles increases, the elementary diagram becomes
less important. In $^{208}$Pb its contribution is of few
parts on a thousand. This fact is in agreement with our
experience of nuclear matter calculations where the
elementary diagrams are irrelevant.

An interesting feature of these calculations is shown in fig.1 where
the optimal correlation functions are shown. The correlation functions
seem to depend only from the interaction and not from the mean--field.
The correlation functions obtained with the $S3$ interaction have
deeper minima than those obtained with the $B1$ interaction.

In fig.2 we compare the uncorrelated proton distribution (full
lines) with those calculated with our theory using the $B1$ and $S3$
interactions (dashed and dashed dotted lines respectively). The effect
of the correlation on the distributions seems to be larger for the
calculations done with the $B1$ interaction than in the case of the
$S3$ interaction. It is remarkable the small effect shown on the
$^{208}$Pb distribution. We found analogous effects on the neutrons
distributions.

In fig.3 we show the proton momentum distributions. In this figure
the presence of short--range correlations shows up at high momentum
value where the FHNC results are orders of magnitude bigger than the
uncorrelated results. The differences between the calculations performed
with the two interactions are small if compared with those with the
uncorrelated momentum distribution.

\section{The model}
The CBF-FHNC theory presented above is a good approximation
for the exact solution of the many-body
Schr\"odinger equation.
A CBF basis may also be constructed, to be used in
perturbation theory. This expansion is expected to have a fast convergence,
since much of the non--perturbative physics, related to the short-range
part of the nuclear potential, is already embedded in  the basis itself.

Unfortunately it is unrealistic to expect of
being able to calculate in a short time the two-nucleon emission cross
sections using this theory. The comparison with the
experimental data produced in the next few years by the
electron accelerators will be done through models. This theory is however
a basis and/or a testing ground for these models.

It is in this perspective that we have started to
develop a simplified model of the theory. This model consists essentially
in calculating only the cluster expansion terms with a single dynamical
correlation line.  The nice feature of this model is that, in spite
of the relative simplicity with respect to the full theory, it
satisfies the same set of sum rules of the full calculation.
Details of the model can be found in ref. \cite{co95}.

In fig.4 we compare the proton density distribution obtained with our
model (dotted line), with that obtained performing the
full FHNC calculation (dashed line).
The model is slightly emphasising the effect of the
correlations.

The relative simplicity of the model allows us to include easily the
effect of the state dependent part of the correlation which are shown in
fig.5. In this figure
the full line is the uncorrelated proton distribution and
the other lines have been obtained including the various state dependent
channels. The number from $1$ to $4$ are indicating the inclusion of
each central channel $(1,\nsigma_i\cdot\nsigma_j,
\ntau_i\cdot\ntau_j, \nsigma_i\cdot\nsigma_j
\ntau_i\cdot\ntau_j)$ and the $5$ and $6$ the isoscalar and
isovector tensor channels\cite{wir88}. The lower panel shows the difference
between uncorrelated and correlated distributions.

We have recently extended our model
to describe excited states.
In fig.6 we present the charge form factors for the discrete transitions
$1f_{7/2}-2s{1/2}^{-1}$ in $^{40}$Ca and $^{48}$Ca. The dashed lines show
the results obtained with the uncorrelated shell model, and the full
lines the results obtained with our model.

\section{Conclusions}
Electron scattering experiments have, so far, investigated
the one-body part of the nuclear many-body wave function.
With the
double coincidence experiments we shall start to investigate the two-body
part of the nuclear wave function. The mean--field models, quite
successful in describing the one--body observables, are inadequate to
describe two-body properties of the nuclei. A profitable comparison
between theory and experiment cannot be obtained adding corrections to
mean--field models fixed to reproduce one--body properties. This because
in these models, part of the effects we want to investigate, those of
the correlation, are already taken into account in an average way.

A proper theoretical description of
the nuclear two-body properties can be obtained only if one treats on
the same ground mean-field, correlations and also final state
interactions. The CBF-FHNC theory provides a theoretical framework to
carry on this ambitious program.
The theory is technically extremely involved and we believe it is
necessary to develop simpler models to describe the forthcoming
experimental data. The CBF--FHNC theory can be
used as a starting point or testing ground for these models which
should be developed.


%
\newpage
%
\vskip 1.cm
\begin{center}
\begin{tabular}{|c|rrrr|}
\hline
$^{4}He$  &  $<v_M>$ & $<v_W>$  & $<T>$  & $<E>$   \\
\hline
  $FHNC/0$  & -132.5 &  24.5  & 83.9  & -44.5   \\
  $FHNC-1$  & -125.6 &        &       & -37.7   \\
     $VMC$  & -123.8 &  24.8  &       & -36.4   \\
\hline
$^{16}O$  &  $<v_M>$ & $<v_W>$  & $<T>$  & $<E>$   \\
\hline
  $FHNC/0$  & -421.6 &  -63.3  & 329.8  & -168.2   \\
  $FHNC-1$  & -403.8 &         &        & -150.4   \\
     $VMC$  & -402.6 &  -62.3  & 327.1  & -150.9   \\
\hline
\end{tabular}
\end{center}

{\bf Tab. 1.} Energies per nucleon, in MeV, for the $^{4}He$ and
$^{16}O$ model nuclei.
\vskip 1.0 cm

\vskip 1.cm
\begin{center}
\begin{tabular}{|c|rrrr|}
\hline
$^{12}C$  &  F1    & F2        &  F3  & F4   \\
\hline
     $V$  & -21.37 & -21.22  & -21.22  & -21.03   \\
   $V_C$  &        &         &   0.64  &   0.63   \\
     $T$  & 19.23  &  19.23  &  19.23  &  19.04   \\
     $E$  & -2.14  &  -1.99  &  -1.35  &  -1.36   \\
\hline
$^{16}O$  &  F1    & F2   & F3  &  F4   \\
\hline
     $V$  & -26.08 &   & -26.08 & -25.74   \\
   $V_C$  &        &   &   0.87 &   0.86   \\
     $T$  & 20.09  &   &  20.09 &  19.80   \\
     $E$  & -5.99  &   &  -5.12 &  -5.08   \\
\hline
$^{40}Ca$  &  F1    & F2   & F3  &  F4   \\
\hline
     $V$  & -32.47  &   & -32.47  & -31.83   \\
   $V_C$  &         &   &   1.95  &   1.91   \\
     $T$  &  23.95  &   &  23.95  &  23.43   \\
     $E$  & -8.52   &   & -6.57   &  -6.49   \\
\hline
$^{48}Ca$  &  F1    & F2   & F3  &  F4   \\
\hline
     $V$  & -33.44  & -33.40 & -33.40  & -32.60   \\
   $V_C$  &         &        &   1.62  &   1.55   \\
     $T$  &  26.11  &  26.11 &  26.11  &  25.49   \\
     $E$  & -7.33   & -7.29  &  -5.67  &  -5.54   \\
\hline
$^{208}Pb$  &  F1    & F2   & F3  &  F4   \\
\hline
     $V$  & -33.86  & -33.84  & -33.84  & -32.96 \\
   $V_C$  &         &         &   3.97  &   3.83   \\
     $T$  &  24.69  & 24.69   &  24.69  &  24.14  \\
     $E$  & -9.17   & -9.15   &   -5.18 &  -4.98   \\
\hline
\end{tabular}
\end{center}

{\bf Tab. 2.} Energies per nucleon, in MeV, for the five nuclei
considered.
\vskip 1. cm

\vskip 1.cm
\begin{center}
\begin{tabular}{|c|rrrrr|}
\hline
B1 & $^{12}C$ & $^{16}O$ & $^{40}Ca$ & $^{48}Ca$ &
$^{208}Pb$   \\  \hline
     $V$  & -24.00 & -27.30  & -33.20  & -32.10 & -34.283 \\
   $V_C$  &   0.67 &   0.86  &   1.90  &   1.56 &   3.819 \\
     $T$  &  19.69 &  17.95  &  20.90  &  21.18 &  20.862 \\
     $E$  &  -3.64 &  -8.49  & -10.40  &  -9.36 &  -9.602 \\
   $E/4$  &  -3.18 &  -7.93  &  -9.74  &  -8.84 &  -9.586 \\
\hline
S3 & $^{12}C$ & $^{16}O$ & $^{40}Ca$ & $^{48}Ca$ &
$^{208}Pb$   \\  \hline
     $V$  & -24.18 & -26.53  & -32.37  & -31.13 & -31.360 \\
   $V_C$  &   0.68 &   0.88  &   1.95  &   1.59 &   3.824 \\
     $T$  &  22.34 &  20.69  &  24.14  &  24.18 &  22.673 \\
     $E$  &  -1.16 &  -4.96  &  -6.28  &  -5.36 &  -4.863 \\
   $E/4$  &  -1.04 &  -4.83  &  -6.17  &  -5.51 &  -4.856 \\
\hline
\end{tabular}
\end{center}

{\bf Tab. 3.} Energies per nucleon, in MeV, obtained with $B1$
and $S3$ interaction. The results labelled $E$ have been
obtained with a FHNC/0 calculation, while those labelled
$E/4$ have been obtained including the elementary diagram.
\vskip 1. cm
\newpage
{\bf Figure captions}
\vskip 1. cm
{\bf Fig. 1.} Correlation functions obtained in the calculations whose
results  are presented in Table 3. \\

{\bf Fig. 2.} Proton density distributions. Full lines, mean--field,
dashed and  dashed--dotted lines, FHNC-1 calculations with $S3$ and $B1$
interaction, respectively. \\


{\bf Fig.3.} Proton momentum distributions. Full lines, mean--field, dashed
and  dashed--dotted lines, FHNC-1 calculations
with $S3$ and $B1$ interaction, respectively.  \\

{\bf Fig. 4.} Proton distribution calculated with the same gaussian
correlation
in FHNC (dashed line) and in the model described in the text (dotted line).
The full line represents the mean--field density distribution.\\

{\bf Fig. 5.}  Proton distributions obtained with our model using state
dependent  correlations taken from nuclear matter\cite{wir88}.
The lower panel show
the differences between the uncorrelated density distribution (full line
in the upper panel) and the various correlated density distributions.
The meaning of the various lines is given in the text.\\

{\bf Fig. 6.} Charge form factor for the transition
$(1f_{7/2}-2s_{1/2}^{-1})$
in $^{40}$Ca (left panel) and $^{48}$Ca (right panel). The dashed lines
represent the results obtained with the uncorrelated shell model
while the full lines are showing the results obtained with our model.

%
\begin{figure}[p]
\begin{center}
\epsfxsize = 0.8\textwidth
\leavevmode\epsffile{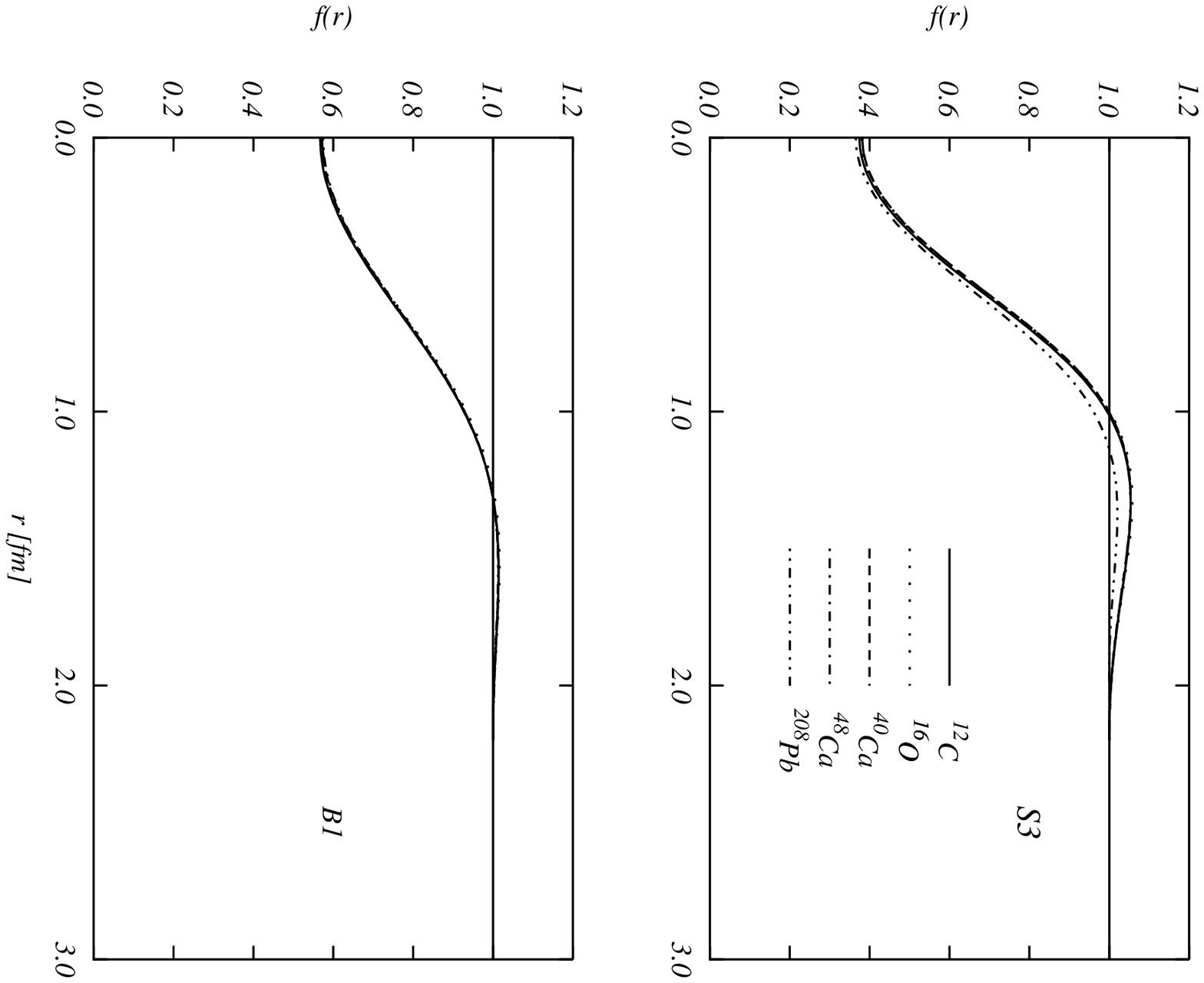}
\end{center}
\caption{
 } \label{}
\end{figure}
\begin{figure}[p]
\begin{center}
\epsfxsize = 0.7\textwidth
\leavevmode\epsffile{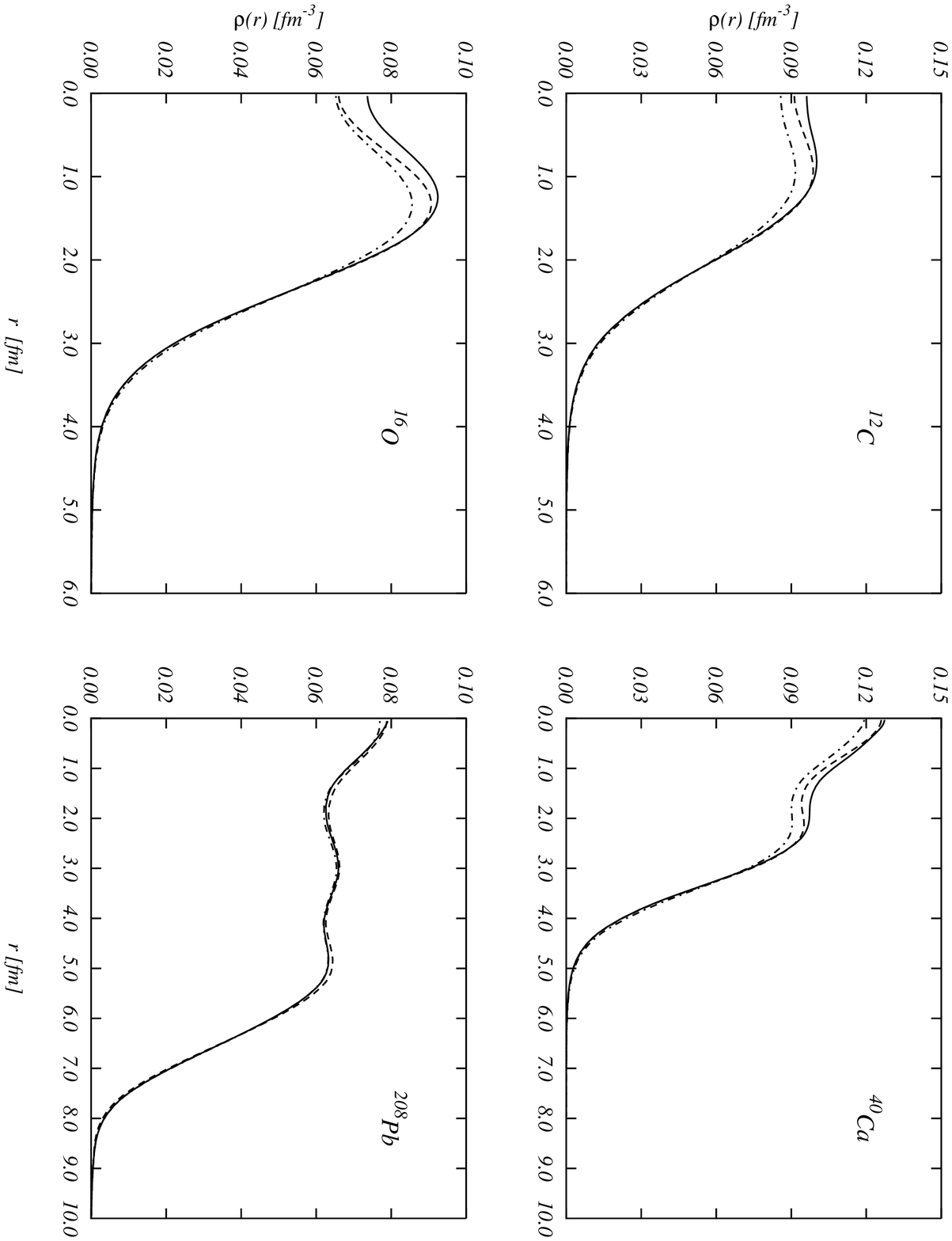}
\end{center}
\caption{
 } \label{}
\end{figure}
\begin{figure}[p]
\begin{center}
\epsfxsize = 0.7\textwidth
\leavevmode\epsffile{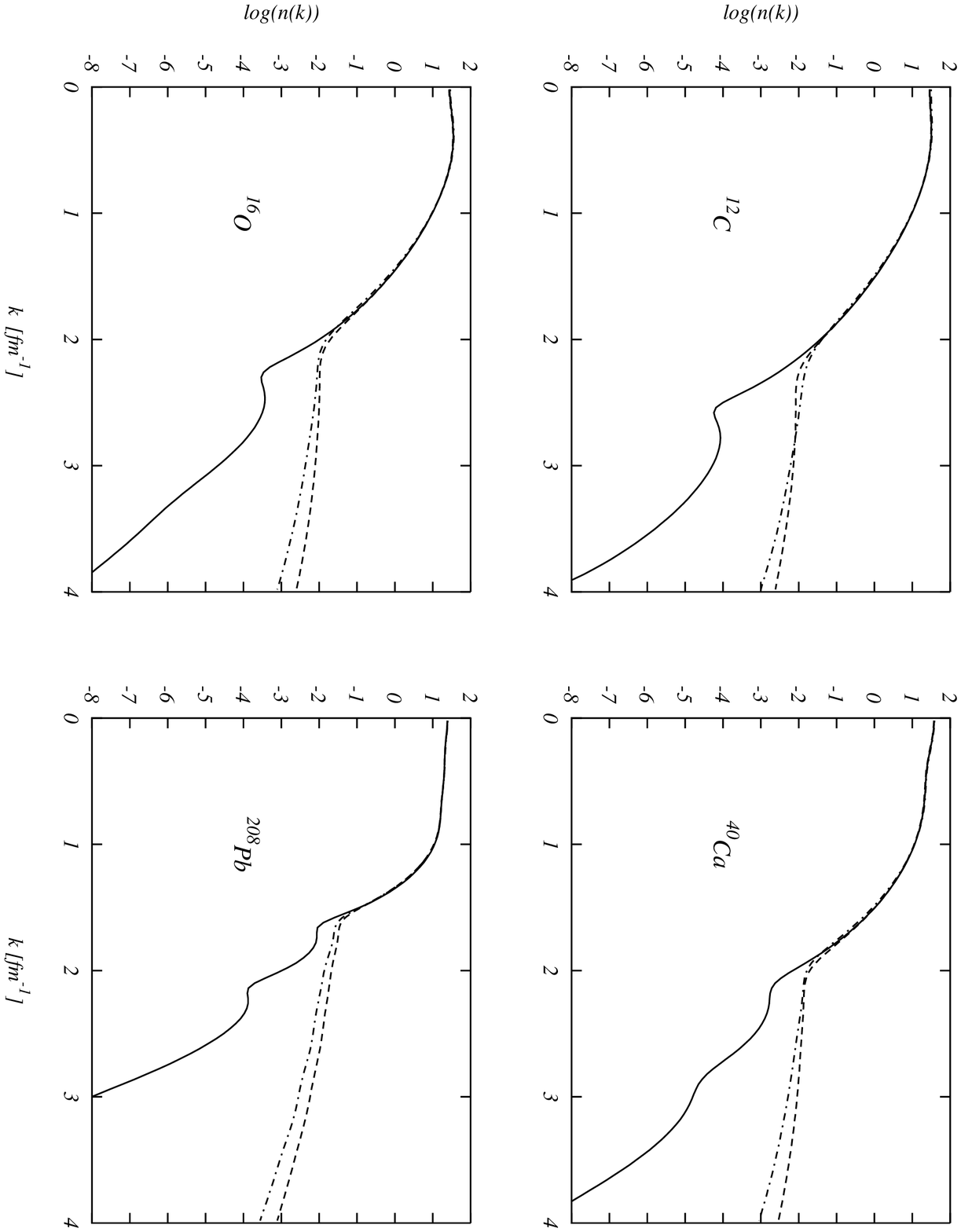}
\end{center}
\caption{
 } \label{}
\end{figure}
\begin{figure}[p]
\begin{center}
\epsfxsize = 0.6\textwidth
\leavevmode\epsffile{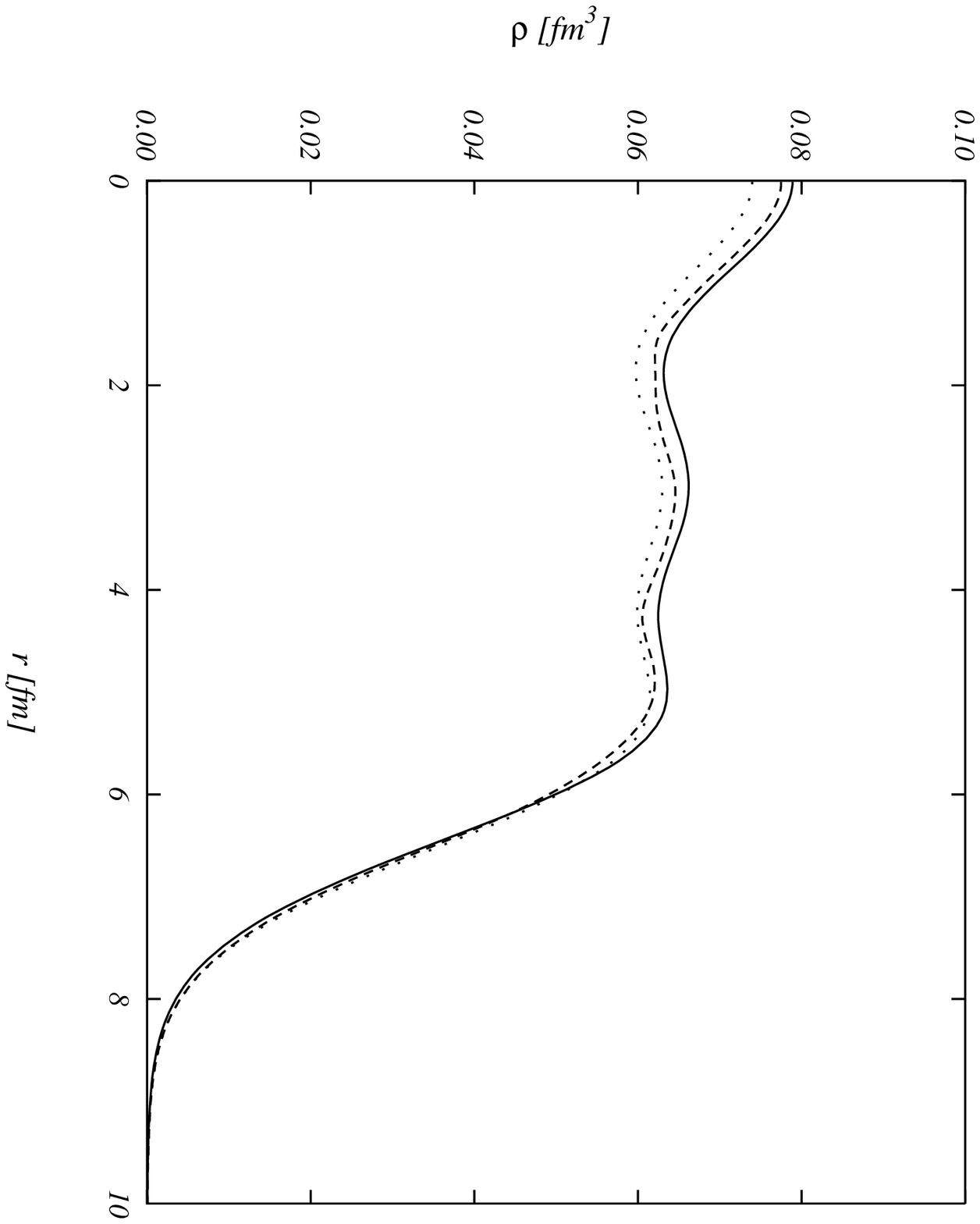}
\end{center}
\caption{
 } \label{}
\end{figure}
\begin{figure}[p]
\begin{center}
\epsfxsize = 0.8\textwidth
\leavevmode\epsffile{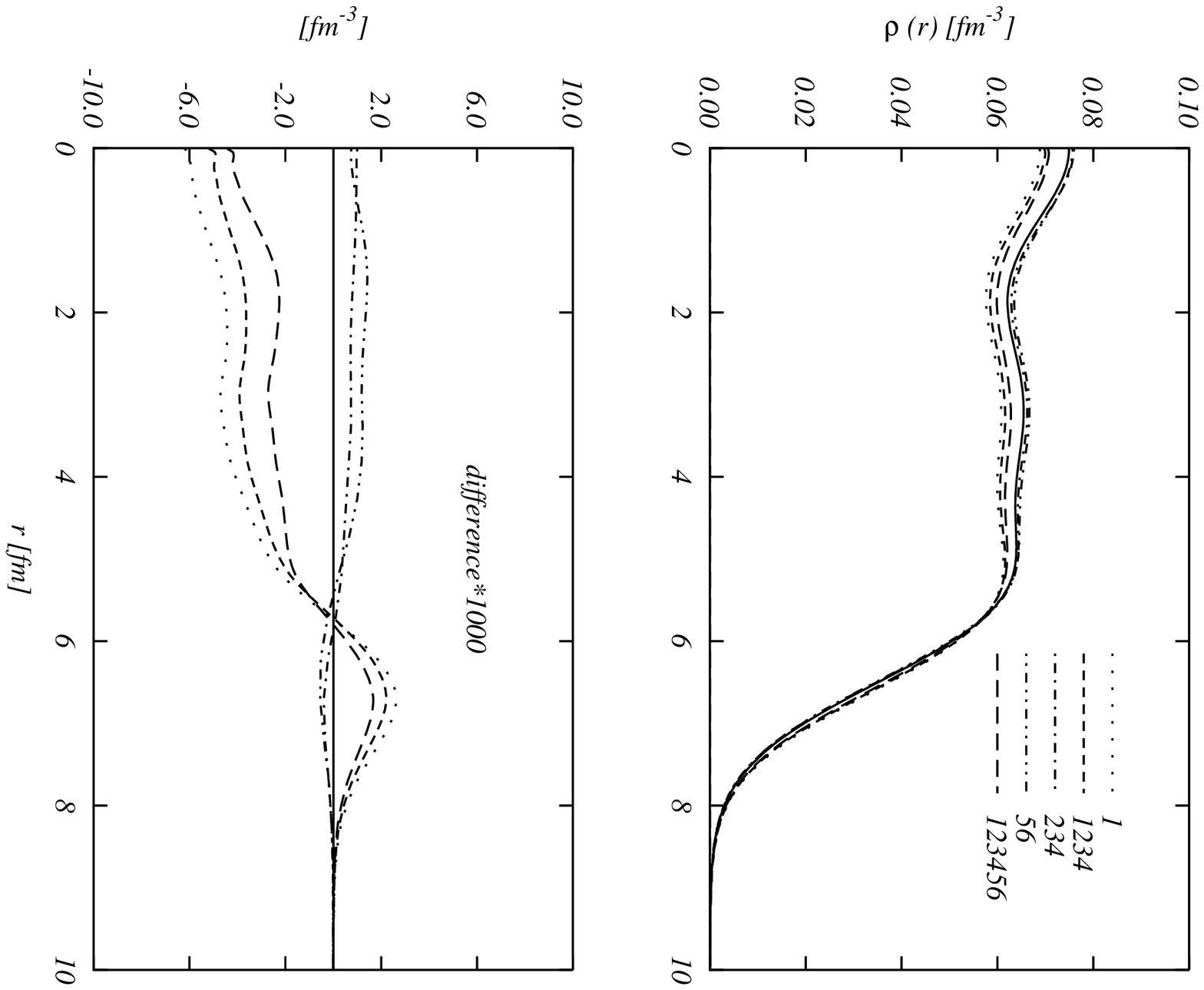}
\end{center}
\caption{
 } \label{}
\end{figure}
\begin{figure}[p]
\begin{center}
\epsfxsize = 0.8\textwidth
\leavevmode\epsffile{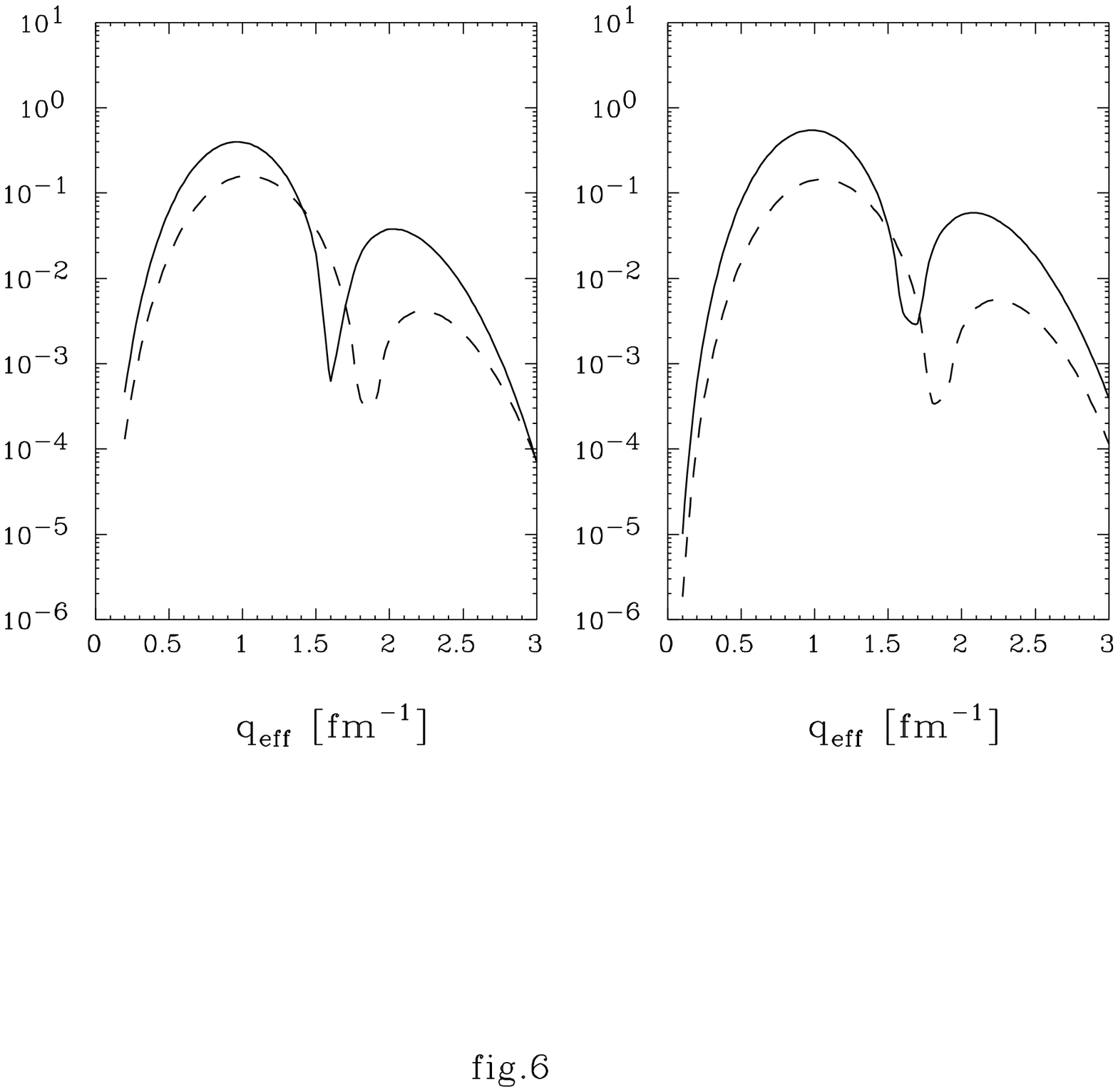}
\end{center}
\caption{
 } \label{}
\end{figure}
%

\end{document}